\documentclass[
 reprint,
 aps, pra,
 amsmath,amssymb,
 11pt,
 final,
tightenlines,
 twoside,
 twocolumn,
 nofloats,
 nobibnotes,
 nofootinbib,
 superscriptaddress,
 noshowpacs,
 showkeys,
 showkeywords,
 centertags]
{revtex4-2}

\usepackage[T2A]{fontenc}
\usepackage[utf8x]{inputenc}
\usepackage[russian,english]{babel}
\usepackage{graphicx}
\usepackage{dcolumn}
\usepackage{bm}
\usepackage{longtable}

\input{maik.rty}

\setcitestyle{authoryear,round}
\def\saoname{Special Astrophysical Observatory,  Russian Academy of Sciences,
              Nizhnii Arkhyz, 369167 Russia}

\def\squareforqed{\hbox{\rlap{$\sqcap$}$\sqcup$}}

\def\sq{\ifmmode\squareforqed\else{\unskip\nobreak\hfil
\penalty50\hskip1em\null\nobreak\hfil\squareforqed
\parfillskip=0pt\finalhyphendemerits=0\endgraf}\fi}

\def\degr{\hbox{$^\circ$}}

\def\arcmin{\hbox{$^\prime$}}

\def\arcsec{\hbox{$^{\prime\prime}$}}

\def\utw{\smash{\rlap{\lower5pt\hbox{$\sim$}}}}

\def\udtw{\smash{\rlap{\lower6pt\hbox{$\approx$}}}}

\def\diameter{{\ifmmode\mathchoice
{\ooalign{\hfil\hbox{$\displaystyle/$}\hfil\crcr
{\hbox{$\displaystyle\mathchar"20D$}}}}
{\ooalign{\hfil\hbox{$\textstyle/$}\hfil\crcr
{\hbox{$\textstyle\mathchar"20D$}}}}
{\ooalign{\hfil\hbox{$\scriptstyle/$}\hfil\crcr
{\hbox{$\scriptstyle\mathchar"20D$}}}}
{\ooalign{\hfil\hbox{$\scriptscriptstyle/$}\hfil\crcr
{\hbox{$\scriptscriptstyle\mathchar"20D$}}}}
\else{\ooalign{\hfil/\hfil\crcr\mathhexbox20D}}%
\fi}}

\newcommand{\ab}{Astrophysical Bulletin }

\newcommand{\aaa}{Astron. and Astrophys. }

\newcommand{\aj}{Astron.~J. }
\renewcommand{\apj}{Astrophys.~J. }
\newcommand{\apjs}{Astrophys.~J. Suppl. }

\newcommand{\azh}{Astron.~Zh. }

\newcommand{\mnras}{Monthly Notices Royal Astron. Soc. }

\begin{document}

\selectlanguage{english}


\title{Early-Type (E, S0) Galaxies in the Catalog of Isolated Galaxies (KIG)}

\author{\firstname{V.~E.}~\surname{Karachentseva}}
 \email{valkarach@gmail.com}
 \affiliation{Main Astronomical Observatory, National Academy of Sciences of Ukraine,  Kyiv, 03143 Ukraine}

\author{\firstname{I.~D.}~\surname{Karachentsev}}
 \affiliation{\saoname}

\author{\firstname{O.~V.}~\surname{Melnyk}}
 \affiliation{Main Astronomical Observatory, National Academy of Sciences of Ukraine, Kyiv, 03143 Ukraine}

\received{October 19, 2020} \revised{January 11, 2021 } \accepted{January 11, 2021}

\begin{abstract}
We use the data of modern digital sky surveys  (PanSTARRS-1, SDSS)
combined with  H\,I-line and far ultraviolet  (GALEX) surveys to
reclassify 165 early-type galaxies from the Catalog of Isolated
Galaxies (KIG). As a result, the number of E- and S0-type galaxies reduced to
91. Our search for companions of early-type KIG galaxies revealed
90 companions around 45 host galaxies with line-of-sight velocity differences
$|dV| < 500$~km\,s$^{-1}$ and linear projected separations \mbox
{$R_{p} < 750$}~kpc. We found no appreciable differences in either
integrated luminosity or color of galaxies associated with the
presence or absence of close neighbors. We found a characteristic
orbital mass-to-luminosity ratio for 26 systems ``KIG galaxy--companion''
to be $M_{\odot}/L_{K} = (74\pm26)
M_{\odot}/L_{\odot}$, which is consistent with the $M_{\rm
orb}/L_{K}$ estimates for early-type isolated galaxies in the
2MIG catalog ($63 M_{\odot}/L_{\odot}$), and also with the $M_{\rm
orb}/L_{K}$ estimates for  E- and S0-type galaxies in the Local
Volume: $38\pm22$ (NGC\,3115), \mbox {$82\pm26$} (NGC\,5128),
\mbox {$65\pm20$} (NGC\,4594). The high halo-to-stellar mass ratio
for  E- and S0-type galaxies compared to the average  $(20\pm3)
M_{\odot}/L_{\odot}$ ratio for bulgeless spiral galaxies is
indicative of a significant difference between the dynamic
evolution of early- and later-type galaxies.
\end{abstract}

\maketitle

\section{INTRODUCTION}

According to generally accepted concepts, early-type (elliptical and
lenticular) galaxies reside mostly in clusters of galaxies,
whereas spiral galaxies are located in the general field and at the
periphery of clusters. This is the well-known
``morphology--density'' effect \citep{oem1974,dre1980}, which
triggered various hypotheses about the origin and subsequent
evolution of early-type galaxies. It is believed that in clusters
of galaxies, where mass density is sufficiently high, early-type
galaxies formed as a result of various\ processes, such as
sweeping-out of gas (ram pressure) dynamical friction, tidal
effects (tidals), merging, etc. \citep{lar2016}. Identification
and analysis of the properties of isolated early-type galaxies as
objects residing in regions with low mass density supposed to be
free of the influence of close neighboring galaxies of
approximately the same luminosity (size) is of special interest.

Many authors have been identifying isolated galaxies both down to
their certain limiting magnitude (or angular size) or within a volume
of fixed distance,  based on the data from
available surveys and catalogs. We published the first Catalog of
Isolated Galaxies  \citep[hereafter referred to as KIG,][]{kara1973}.
Isolated objects among almost  30\,000 galaxies of the Zwicky
catalog \citep{Zwi1968} with apparent magnitudes $m\leq15.7$ and
declinations $\delta>-2\degr30\arcmin$ were identified by uniformly applying
the isolation criterion to all galaxies of the  POSS-I
photographic sky survey. The criterion takes into account fore-
and background objects, namely: isolated galaxies were considered to be
those with such angular diameter  $a_{i}$ that  their
``neighbors'' with diameters  \mbox {$1/4a_{i} < a_{j}< 4 a_{i}$}
were located at projected separations $R_{ij}\geq 20 a_j$. Of 1050
KIG galaxies about  16\% are early-type systems  (E, S0), whereas
the remaining ones are spiral and irregular galaxies and galaxies
of unclear type.

Given the typical size of about 20~kpc, according the selection
criteria a KIG galaxy should have no ``significant'' neighbors
(i.e., those that influence its dynamic isolation) within the
volume of $2\times10^{8}$~kpc$^{3}$ \citep{kara1980}.
\citet{ada1980} showed that KIG galaxies should not have been
influenced by neighboring galaxies over the past several billion
years, and hence they must have been isolated throughout almost
their entire lifetime. \citet{ver2007a, ver2007b} applied
statistical criteria (based on local density and tidal force) to
assess the degree of isolation and showed that the evolution of
KIG galaxies was driven by internal processes \citep{ada1980,
ver2007a, ver2007b}.

\citet{ada1980} reclassified  165 presumed  E- and S0-type KIG
galaxies, confirming 120 of them as early type galaxies (ETGs).
 \citet{sto2004} performed a detailed analysis of the KIG-sample and
found 65 isolated elliptical and 37 isolated S0-type galaxies, i.e., according
to their data, the KIG contains about  9.7\% ETGs. \citet{sul2006}
used the POSS-II photographic sky survey for their new visual
classification of KIG galaxies and found the fraction of
early-type galaxies in KIG to be of about 14\%. \citet{her2008}
classified  579 KIG galaxies using  SDSS\,DR6 data and the CAS
system \citep{con2003}. They found the fraction of  E+S0 galaxies
to be significantly smaller---8.5\% (3.5+5\%)---than that
obtained by \citet{sul2006}. \citet{but2019} reported a
new classification of 719 KIG galaxies and found early-type
systems to make up  14\% (5.3\% and 8.7\% for E and S0 galaxies,
respectively) of the sample.

The AMIGA project\footnote{http//www.iaa.es/AMIGA.html}  team made
a very important contribution to the study of the properties of
KIG galaxies (see also \citet{sul2010}).

New sky surveys were released since the publication of the KIG
catalog---SDSS \citep{yor2000}, 2MASS \citep{skr2006}, and 2MXSC
\citep{jar2000}. They were used in compiling new catalogs and
lists of isolated galaxies: UNAM-KIAS \citep{her2010} based on
SDSS\,DR6; 2MIG \citep{kara2010} based on the 2MASS  infrared
all-sky survey; the  LOG catalog \citep{kar2011} of isolated
galaxies in the Local Supercluster volume; \citep{arg2015} list
based on SDSS\,DR10 \citep{ahn2014}, and others. Some properties of
isolated galaxies were described, in particular, in
\citet{fer2012, fer2013, lar2016, lar2018}.

While compiling catalogs and lists of isolated galaxies, the above
authors used various modifications of the KIG isolation
criterion. They adopted different values for the allowed magnitude
difference $dm$ between the isolated galaxy and its possible
neighbors, allowed radial velocity difference  $dV$, and their
projected separation $R_{p}$. These characteristics vary over rather
wide ranges (\mbox {$dm = 1$--$3$}~mag,
$dV=300$--$1000$~km\,s$^{-1}$, $R_{p} = 250$--$1000$~kpc), see,
e.g., \citet{red2004, her2010, arg2015}. The most stringent
criterion for galaxies believed to be isolated is described in
\citet{mar2004}: $|dV|$= 350~km\,s$^{-1}$, $R_{p}$=2500~kpc, and
the absence of nearby companion brighter than \mbox
{$M_{V}=-16.5$}. This criterion revealed only nine KIG galaxies;
the authors performed $BVR$ photometry of these galaxies,
determined their types, and tried to find companions even without
the knowledge of radial velocities. The small number of galaxies
considered makes it impossible for us to make any comparisons.

Identification of isolated galaxies in new catalogs goes along
with their morphological classification. Note that morphological
classification of galaxies even now remains to a great extent
subjective. This classification, which began with the works of
Hubble, de Vaucouleurs and Sandage, is continued in
\citet{but2019} (see references therein) and \citet{gra2019},
where Graham provides an extensive review of studies dedicated the
classification of galaxies. We return to this issue in Section~2.

According to the classical definition of elliptical galaxies, they
can be described as smooth, regular-shaped galaxies without dust
or gas and without structural details in the center and ''body``
of the galaxy. They have red colors and, usually, absorption-line
spectrum. As for lenticular galaxies, Hubble back then believed
them to be intermediate between elliptical and spiral galaxies.

In this paper we adhere to the above cutoff values of parameters,
especially, given that available observational data allow one to
quite definitively classify elliptical and lenticular KIG
galaxies. We considered only early-type galaxies tagged in the KIG
as being of E or S0  (or E--S0) type. The authors of recent
studies subdivide lenticular galaxies into two classes: (1)---pure-bulge galaxies
and (2) galaxies with disk properties: bluer
color, emission lines in the spectrum, etc. (see \citet{fra2018,
tou2020} and references therein). A description of the properties
of lenticular galaxies can also be found in the extensive
introduction to paper \citet{dee2020}. Its authors propose, based
on the data of the SAMI survey \citep{gre2018}, two possible
scenarios for the formation of S0-type galaxies: either fading of
spirals or formation as a result of galaxy mergers. The results of
the photometry of 42 galaxies are reported in \citet{sil2020}; the
above authors point out a probably different dynamic history of
S0-type galaxies in different environments.

Here, we use modern sky
surveys to perform a new classification of early-type galaxies
(ETG) from the KIG catalog for two reasons: (1)~earlier
classifications reported in 1973 and 2006 have become outdated and
(2) the other catalogs of isolated galaxies mentioned above are
based on different sky-survey data---2MASX and SDSS. We
introduce a new classification and subdivide KIG galaxies into
ETGs without companions and ETGs with insignificant (small) companions---we use the
latter to compute orbital masses of the ``ETG galaxy--companion'' systems.

\renewcommand{\baselinestretch}{0.9}
\begin{table*}
\caption{Early-type KIG galaxies without companions. (1)---galaxy name,
(2)--- compactness according to the catalog of
Zwicky et al.: compact---c, very compact---vc, extremely
compact---ec, (3)---the type according to  HyperLEDA, (4)---the type estimated based on PanSTARRS-1}
\medskip
\begin{tabular}{c|c|c|c||c|c|c|c}
\hline
KIG &   Zwicky  &   Type (LEDA)  &   Type (PS-1) & KIG &   Zwicky  &   Type (LEDA)  &   Type (PS-1)\\
\hline
(1) & (2)& (3)& (4) & (1)& (2)& (3)& (4) \\
\hline
14  &    & S0       & S0       & 636 &    & S0   & S0 pec \\
57  &    & E-S0, S2 & S0       & 670 & vc & E-S0 & S0     \\
99  &    & S0-a     & S0       & 684 &    & E    & E      \\
101 &    & E-S0     & E-S0     & 701 &    & E?   & S0 pec \\
110 &    & E        & E        & 763 &    & E    & S0     \\
118 &    & E-S0     & E-S0     & 770 & vc & E    & E      \\
127 &    & E-S0     & E        & 792 & c  & S0   & E      \\
136 &    & E        & E        & 816 &    & S0   & S0     \\
174 & c  & S?       & E        & 820 &    & E-S0 & S0     \\
179 & c  & E-S0     & S0       & 823 & c  & E?   & E      \\
256 & ec & E-S0     & E        & 824 &    & E    & S0     \\
378 &    & E-S0     & E pec    & 826 & c  & E    & E      \\
387 & c  & E-S0     & E        & 827 & vc & E-S0 & E      \\
412 & c  & E        & E        & 833 & ec & E-S0 & E      \\
443 &    & S0-a     & S0       & 836 & vc & E    & E      \\
452 & c  & E-S0     & S0       & 845 & c  & E    & S0     \\
462 & c  & E-S0     & S0 pec   & 865 & c  & E-S0 & E      \\
490 & c  & S0, ring & S0 pec   & 870 &    & E-S0 & E pec  \\
521 &    & S0       & S0       & 877 &    & E-S0 & E pec  \\
529 &    & E        & E-S0     & 894 & c  & E-S0 & S0 pec \\
570 &    & S0-a     & S0 pec   & 896 & c  & E-S0 & S0     \\
574 & vc & E        & E        & 920 &    & E-S0 & S0     \\
582 & c  & E        & E-S0 pec & 981 & c  & E    & S0     \\
\hline
\end{tabular}
\end{table*}
\renewcommand{\baselinestretch}{1.0}

The paper has the following layout.
\begin{list}{}{
\setlength\leftmargin{4mm} \setlength\topsep{2mm}
\setlength\parsep{0mm} \setlength\itemsep{2mm} }
 \item Section~2---identification and morphological classification of early-type galaxies in the KIG based on  PanSTARRS-1 survey data.
 \item Section~3---results of the search for companions/neighbors and description of their main properties.
 \item Section~4---comparison of the properties of early-type KIG galaxies with and without their satellites/neighbors.
 \item Section~5---determination of the orbital masses of some KIG galaxies based on the data about their nearest neighbors.
 \item Section~6 presents the concluding remarks.
 \end{list}

\section{MORPHOLOGICAL CLASSIFICATION OF EARLY-TYPE KIG GALAXIES}

In our work we proceeded from the assumption that all 165 galaxies
classified as E or S0 in the KIG are isolated and imposed no
constraints with respect to their radial velocities, apparent
magnitudes, or sky positions. After excluding 74 galaxies
that turned out to be spiral,
we use the measured properties provided by various databases for
the remaining ETG galaxies. We use HyperLEDA \citep{mak2014} as
our source of integrated magnitudes $b_{t}$, Galactic and internal
extinction $A_{G}$ and $A_{i}$, ($A_{i}$=0 for E and S0-type
galaxies), 21-cm-line magnitudes $m_{21}$, and absolute
$B_{t}$-band magnitudes $m_{\rm abs}$. We adopt from NED the
radial velocities $V_{\rm LG}$ (km\,s$^{-1}$) in the frame of the
centroid of the Local group, compute the radial velocity
differences, and the linear  ``KIG galaxy--companion''
separations. We compute the distances and absolute properties of
the galaxies from their $V_{\rm LG}$ adopting the Hubble constant
of  $H_{0}$ = 73~km\,s$^{-1}$\,Mpc$^{-1}$. We determine the
$g$--$r$ and $g$--$i$ colors from SDSS survey data in the
close-to-AB magnitude system
\footnote{http://classic.sdss.org/dr7/algorithms/fluxcal.html\#sdss2ab}.

We adopt the far=ultraviolet magnitudes $m_{FUV}$ from the GALEX
survey \citep{mar2005}. To estimate the stellar masses of  E- and
S0-type galaxies from their  $K$-band luminosities, in Section~6
we use the directly measured  $K_{s}$-band magnitudes adopted from
the NED database. We determine the $K$-band magnitudes for
companions of various morphological types from their $B$-band
magnitudes and morphological type $T$ via relation $$\langle
B-K\rangle_{\rm corr} = 4.60 - 0.25\times T,$$ because the
$K$-band magnitudes of late-type galaxies are highly
underestimated in the 2MASS survey. We compute the integrated
star-formation rate $SFR$ for all galaxies by formula~(6) from
\citet{mel2017} with extinction corrections applied (formulas~(2)
and~(5) in the same paper).

Our classification is based on PanSTARRS-1 (PS-1) sky survey
\citep{cha2016}. We estimate the galaxy types mostly based on the
shape of the object, but also take into account the presence of
21-cm H\,I lines, bright optical emission lines, and emissions in the far
ultraviolet ($FUV$) according to GALEX data \citep{mar2005}

We present the results of classification in Table~1 (KIG galaxies
without satellites) and Table~2 (galaxies with
satellites/neighbors). We consider satellites to be galaxies  that
are more than 1~mag fainter than the ``host'' KIG galaxy. The
apparent magnitudes of neighbors are approximately comparable to
those of KIG galaxies. Where necessary, we consider satellites and
neighbors separately. Note that only one galaxy -- KIG\,664 (S0
according to our estimate) -- has no measured radial velocity and
we therefore could not include it into either Table~1 or Table~2.

Typical elliptical galaxies that we classified by their shape
have absorption-line spectra and exhibit no emissions either in
the optical bands or in $FUV$. The properties of S0 galaxies
were described, in particular, in the Introduction to
paper~\citet{dee2020}.

\citet{ash2019} describe the selection criterion as
well as optical and  H\,I properties of the galaxies that they
believed to be extremely isolated, IEG sample, $N=25$. These
galaxies have absolute $B$- magnitudes in the [-14.2; -20.7] range
and $\langle B-V\rangle$ = 0.58. The same criterion was used to
select extremely isolated early-type galaxies (IEG) in the  SDSS
survey (\citet{fus2012}, see also the classification there). The
above authors write about careful selection, which left only 33
galaxies. However, 14 of these are dwarf galaxies being 1--2~magnitude
fainter than typical early-type galaxies. We checked E-type
galaxies from Table~1 in \citet{fus2012} and found them to be
characterized by bright emission lines typical for BCD galaxies.
According to our definition, they are not classical elliptical
galaxies, although they, on the average, have a round shape. The
differences between our data and sample \citet{fus2012} may be due
to a selection effect typical for flux-limited surveys. Therefore because
of different depths of the samples (200 and 70~Mpc for our sample
and that of \citet{fus2012}, respectively), the SDSS sample is
shifted toward blue galaxies of low luminosity located at small redshifts.

\renewcommand{\baselinestretch}{0.8}
\onecolumngrid
\begin{longtable}{l|c|c|c}
\caption {Early-type KIG galaxies with satellites/neighbors.
(1)---galaxy name, (2)---compactness according to the
catalog of Zwicky et al.: compact---c, very compact---vc,
extremely compact---ec,
(3)---type according to HyperLEDA, (4)---type estimated from PanSTARRS-1}\label{1}\\
 \hline
\multicolumn{1}{c|}{Galaxy}  &   Zw  &   T (LEDA) &   T (PS-1)\\
\hline
\multicolumn{1}{c|}{(1)} &   (2) &   (3) &   (4) \\
\hline
\endfirsthead
\caption{(Continued) }\\
\hline
\multicolumn{1}{c|}{Galaxy}  &   Zw  &   T (LEDA) &   T (PS-1)\\
\hline
\multicolumn{1}{c|}{(1)} &   (2) &   (3) &   (4) \\
\hline
\endhead

\hline
\endfoot

\hline
\endlastfoot

KIG\,24                           &    & E       & S0      \\
neighb.1, CGCG\,409-21            &    & S0-a    & S0      \\
KIG\,25                           &    & S0      & S0      \\
sat.1, UGC\,287                   &    & Scd     & Scd     \\
KIG\,79                           &    & E-S0    & S0      \\
neighb.1, CGCG\,461-14            &    & S0      & S0      \\
neighb.2, UGC\,1485               &    & Sc      & Sc pec  \\
neighb.3, CGCG\,461-20            &    & Sc      & S pec   \\
KIG\,89                           &    & E       & E       \\
sat.1, KKH\,8                     &    & Ir      & Ir      \\
KIG\,111                          & c  & E       & E       \\
sat.1, AGC\,122418                &    & G       & Sm      \\
KIG\,161                          &    & Sa      & S0      \\
sat.1, PGC\,138829                &    & G       & Scd     \\
KIG\,184                          &    & SABa    & S0      \\
neighb.1, CGCG\,234-15            &    & SABb    & Sb      \\
KIG\,189                          &    & E       & E       \\
sat.1, PGC\,2228154               &    & G       & Ir      \\
sat.2, SDSS\,J072524.12+422559.1  &    & S?      & Im      \\
KIG\,228                          & c  & E       & E       \\
sat.1, WISEA\,J080731.37+555342.1 &    & G       & BCD     \\
KIG\,233                          &    & E-S0    & S0      \\
sat.1, WISEA\,J081051.83+273404.2 &    & G       & BCD     \\
neighb.1, AGC\,183054             &    & S?      & Sd      \\
KIG\,245                          &    & E       & E       \\
sat.1, AGC\,181571                &    & Sm      & Ir      \\
sat.2, AGC\,188871                &    & G       & Sc      \\
KIG\,264                          & c  & S0-a    & S0      \\
sat.1, KUG\,0832+305              &    & S?      & Scd     \\
neighb.1, Mrk\,390                &    & Sc      & Sb pec  \\
KIG\,303                          &    & S0      & S0      \\
sat.1, AGC\,193009                &    & E       & S0 pec  \\
sat.2, AGC\,191082                &    & SBc     & Scd     \\
sat.3, SDSS\,J090703.40+034905.7  &    & SBc     & Scd     \\
KIG\,380                          &    & E       & E       \\
sat.1, AGC\,731423                &    & S?      & BCD     \\
KIG\,396                          &    & E-S0    & E       \\
sat.1, SDSS\,J100413.44+602214.1  &    & Sd      & Sd      \\
sat.2, KUG\,0958+599              &    & Sd      & BCD     \\
neighb.1, UGC\,5408               &    & E-S0    & BCD     \\
neighb.2, CGCG\,289-27            &    & E-S0    & S0      \\
KIG\,413                          &    & S0-a    & S0      \\
sat.1, PGC\,1188869               &    & S0      & S0      \\
sat.2, AGC\,204701                &    & S?      & Im      \\
sat.3, AGC\,204919                &    & Scd     & Sc      \\
sat.4, AGC\,204920                &    & Sm      & Sm      \\
sat.5, PGC\,1181655               &    & E       & BCD     \\
neighb.1, AGC\,201427             &    & Sa      & Sa pec  \\
KIG\,415                          & vc & E       & S0      \\
sat.1, AGC\,203492                &    & S?      & Sc      \\
KIG\,425                          & c  & E       & E       \\
sat.1, PGC\,2628623               &    & Sd      & Sm      \\
KIG\,426                          & vc & E-S0    & S0      \\
sat.1, PC\,1034+4938              &    & emis.g. & BCD     \\
sat.2, PGC\,2336611               &    & E       & S0      \\
sat.3, PGC\,2346694               &    & Sc      & Spec    \\
sat.4, PGC\,2335306               &    & E       & E       \\
KIG\,437                          &    & E       & E-S0    \\
sat.1, MCG\,9-18-17               &    & E       & S0      \\
sat.2, WISEA\,J104402.83+523034.7 &    & S ?     & Sc?     \\
KIG\,480                          &    & Sab     & S0      \\
sat.1, AGC\,217484                &    & Sm      & Sdm     \\
neighb.1, UGC\,6437               &    & Sbc     & Sc      \\
neighb.2, AGC\,12238              &    & SBbc    & Sbc     \\
KIG\,513                          & vc & E       & E       \\
sat.1, AGC\,719642                &    & S?      & Sm      \\
neighb.1, AGC\,719646             &    & Sbc     & Sbc     \\
KIG\,517                          & c  & S0      & S0      \\
sat.1, WISEA\,J120240.67+261248.9 &    & G       & Sd      \\
sat.2, WISEA\,J120344.73+260345.8 &    & E       & S0      \\
KIG\,557                          & c  & E       & E       \\
sat.1, PGC\,1162105               &    & E       & E       \\
sat.2, PGC\,1161248               &    & S0      & S0      \\
sat.3, PGC\,3298012               &    & S?      & Sbc     \\
sat.4, PGC\,1157914               &    & S?      & Sc      \\
sat.5, PGC\,3297967               &    & G       & Sbc     \\
KIG\,578                          & c  & E       & E       \\
sat.1, WISEA J131629.64+200518.5  &    & S?      & BCD     \\
sat.2, WISEA J131728.70+200130.2  &    & G       & S?      \\
KIG\,595                          &    & E       & E       \\
sat.1, WISEA J133911.75+612916.0  &    & E?      & S0      \\
sat.2, PGC\,2619551               &    & S?      & S0      \\
KIG\,596                          &    & S0-a    & S0 pec  \\
sat.1, PGC\,2625488               &    & Sc      & BCD     \\
KIG\,599                          &    & S0      & S0 pec  \\
sat.1, PGC\,2097287               &    & S?      & Sdm     \\
KIG\,602                          &    & S?      & S0      \\
sat.1, PGC\,1681951               &    & G       & Sc?     \\
sat.2, PGC\,1678559               &    & Sbc     & Sc      \\
sat.3, PGC\,1678503               &    & Sb      & S0      \\
sat.4, KUG\,1350+232              &    & Sbc     & Sbc     \\
sat.5,WISEA\,J135409.10+230454.8  &    & S?      & Im      \\
KIG\,614                          &    & Sbc     & S0      \\
sat.1, WISEA\,J141057.79+215317.9 &    & G       & S0      \\
neighb.1, PGC\,1657978            &    & S?      & S0-a    \\
KIG\,623                          & vc & E       & E       \\
sat.1, WISEA\,J141823.99+193432.4 &    & E       & BCD     \\
sat.2, WISEA\,J142021.46+202332.5 &    & G       & Sd      \\
KIG\,685                          & c  & E       & Epec    \\
sat.1, WISEA\,J152927.49+565558.4 &    & SBc     & Sc      \\
KIG\,703                          & ec & E       & E-S0    \\
sat.1, WISEA\,J154723.56+221143.6 &    & G       & BCD     \\
KIG\,705                          & vc & E-S0    & Epec    \\
sat.1, WISEA\,J154720.49+370255.6 &    & S?      & Sm      \\
KIG\,722                          &    & E       & E       \\
sat.1,WISEA\,J160822.82+093957.4  &    & E?      & E-S0    \\
KIG\,732                          & c  & E       & E       \\
sat.1, Mrk\,498                   &    & G       & BCD     \\
KIG\,768                          & vc & E-S0    & S0 pec  \\
sat.1, WISEA\,J164441.66+194636.9 &    & Sbc     & Sc      \\
neighb.1, CGCG\,110-4             &    & Sc      & Scd     \\
KIG\,771                          & c  & E       & E       \\
sat.1, PGC\,1678008               &    & S?      & E       \\
sat.2, WISEA\,J164645.65+225147.1 &    & E?      & S0      \\
sat.3, PGC\,1678062               &    & E       & E       \\
sat.4, WISEA\,J164709.15+225849.6 &    & S?      & Ir      \\
sat.5, WISEA\,J164715.62+224940.9 &    & G       & E       \\
sat.6, WISEA\,J164726.19+225519.5 &    & S?      & E       \\
sat.7, PGC\,1679574               &    & S?      & Sc      \\
sat.8, PGC\,1676423               &    & E       & E-S0    \\
KIG\,898                          &    & E-S0    & E       \\
neighb.1, PGC\,165874             &    & G       & S0-a    \\
KIG\,903                          &    & E       & S0      \\
sat.1, WISEA\,J211516.19+095346.8 &    & S?      & BCD     \\
KIG\,921                          & c  & E-S0    & merger? \\
neighb.1, RFGC\,3770              &    & Sc      & Scd     \\
KIG\,1015                         & c  & E       & S0      \\
neighb.1, NGC\,7628               &    & E       & E pec   \\
KIG\,1025                         & ec & E-S0    & S0 pec  \\
sat.1, AGC\,331187                &    & IAB     & Sm      \\
KIG\,1042                         & vc & E       & E       \\
sat.1, AGC\,331919                &    & G       & Sc      \\
sat.2, AGC\,333425                &    & G       & Sd      \\
KIG\,1045                         &    & E       & S0      \\
sat.1, WISEA\,J235442.78+052254.1 &    & S?      & Sc pec  \\
\end{longtable}
\twocolumngrid
\renewcommand{\baselinestretch}{1.0}

Isolated galaxies marked in the  \citet{Zwi1968} catalog as
``compact'', ``very compact'', or ``extremely compact'', appear in
PS-1 images as normal elliptical and lenticular galaxies. The only
exceptions are KIG\,256, KIG\,705, KIG\,732, KIG\,770, KIG\,826,
and KIG\,833, which appear sufficiently compact even in PS-1. It
is clear that on POSS-I images, which were taken about 60 years
ago, diffuse envelopes of distant galaxies could not be discerned
to say nothing about the structural details of these systems. Our
new classification reduced the  fraction of early-type galaxies
(ETG) in the KIG approximately by half:  91 among 1050 galaxies,
i.e.,  8.7\%, which agrees better with the data from
\citet{her2008}. The number of E-type galaxies is approximately
equal to that of S0-type galaxies, 40 (44\%) and 44 (48\%),
respectively, and the number of  E-S0-type galaxies is 7 (8\%).
The twofold reduction of the fraction of  ETG as a result of about
half of them being reclassified as spirals is due to better
quality of digital CCD images (broader dynamic range) and more
rigorous selection. Our results show that isolated ETG galaxies
are rather numerous and make up an interesting sample for further
study. We compared our data with the morphological classification
of Rampazzo et al. \citep{ram2020} based on deep photometry.
As a result, we excluded  KIG\,481, KIG\,620, KIG\,637, KIG\,644,
KIG\,733, and KIG\,841 from ETG galaxies because they
are classified as bona fide spirals on PS-1 images. The remaining
14 galaxies are early-type objects. The details can be checked in
Tables~1 and~4 in~\citet{ram2020}, as well as in our Tables~1
and~2.

The remaining early-type galaxies in the KIG exhibit morphological
peculiarities in approximately 20\% of the cases. These
peculiarities may be due both to their internal evolution and to
recent merging with fainter objects.

Table~3 lists some of the basic properties  (means and standard
errors of mean) for ETG galaxies in the KIG. The top six rows of the table
describe the characteristics  of the ETG galaxies proper. The four
bottom rows refer to neighbors and satellites of KIG galaxies.

\begin{turnpage}

\begin{table*}
\caption {Some basic properties (means and standard errors of mean) for ETG
galaxies in the KIG}
\medskip
\begin{tabular}{l|c|c|c|c|c|c|c|c|c|c|c|c}
\hline
\multicolumn{1}{c|}{Type (PS-1)} & $N$ & $M_{K_b}^{\rm cor}$ & $\log M^*$  & $M_{\rm abs}^{\rm LEDA}$  & $N$ & $\log(SFR)$ & $\log(sSFR)$ & $N$ & $\log M_{HI}$ & $N$ & $g-r$ & $g-i$\\
\hline
\multicolumn{1}{c|}{(1)}& (2)& (3)& (4)& (5)& (6)& (7)& (8)& (9)& (10)& (11)& (12)& (13) \\
\hline
 E (all)     & $43$ & $-24.15\pm0.13$ & $10.99\pm0.05$ & $-20.28\pm0.14$ & $33$ & $-1.21\pm0.09$ & $-12.17\pm0.07$ & $4 $ & $9.78\pm1.12$  & $25$ & $0.81\pm0.01$ & $1.21\pm0.01$ \\
 S0 (all)    & $48$ & $-24.24\pm0.11$ & $11.02\pm0.04$ & $-20.38\pm0.12$ & $37$ & $-1.19\pm0.07$ & $-12.25\pm0.06$ & $12$ & $9.27\pm0.16$  & $36$ & $0.75\pm0.02$ & $1.10\pm0.05$ \\
 E (no sat)  & $21$ & $-24.08\pm0.19$ & $10.96\pm0.08$ & $-20.19\pm0.21$ & $18$ & $-1.12\pm0.11$ & $-12.04\pm0.12$ & $2 $ & $9.33\pm0.11$  & $5 $ & $0.82\pm0.02$ & $1.22\pm0.03$ \\
 S0 (no sat) & $25$ & $-24.25\pm0.15$ & $11.03\pm0.06$ & $-20.38\pm0.15$ & $22$ & $-1.17\pm0.10$ & $-12.21\pm0.07$ & $6 $ & $9.26\pm0.18$  & $16$ & $0.73\pm0.03$ & $1.17\pm0.10$ \\
 E (sat)     & $22$ & $-24.23\pm0.18$ & $11.02\pm0.07$ & $-20.36\pm0.20$ & $15$ & $-1.32\pm0.13$ & $-12.32\pm0.08$ & $2 $ & $10.23\pm3.10$ & $20$ & $0.80\pm0.01$ & $1.21\pm0.02$ \\
 S0 (sat)    & $23$ & $-24.22\pm0.16$ & $11.02\pm0.06$ & $-20.38\pm0.17$ & $15$ & $-1.22\pm0.09$ & $-12.31\pm0.09$ & $6 $ & $9.28\pm0.25$  & $20$ & $0.76\pm0.02$ & $1.11\pm0.04$ \\
\hline
\multicolumn{13}{l}{$dm<1$} \\
 E           & $4 $ & $-22.68\pm0.68$ & $10.40\pm0.27$ & $-19.45\pm0.51$ & $3 $ & $-1.00\pm0.46$ & $-11.18\pm0.49 $& $2$  & $9.56\pm0.20$  & $3 $ & $0.82\pm0.14$ & $1.16\pm0.17$ \\
 S0          & $16$ & $-23.61\pm0.17$ & $10.77\pm0.07$ & $-20.17\pm0.14$ & $10$ & $-0.54\pm0.17$ & $-11.32\pm-0.21$& $8$  & $9.52\pm0.13$  & $14$ & $0.59\pm0.05$ & $0.91\pm0.07$ \\
\hline
\multicolumn{13}{l}{$dm\geq1$} \\
E            & $40$ & $-20.64\pm0.25$ & $9.58\pm0.10$  & $-17.74\pm0.20$ & $23$ & $-1.18\pm0.11$ & $-10.95\pm0.14$ & $10$ & $9.17\pm0.15$  & $35$ & $0.54\pm0.04$ & $0.80\pm0.05$ \\
S0           & $30$ & $-21.07\pm0.30$ & $9.76\pm0.12$  & $-18.19\pm0.26$ & $20$ & $-1.05\pm0.08$ & $-10.66\pm0.14$ & $10$ & $9.19\pm0.07$  & $26$ & $0.50\pm0.04$ & $0.75\pm0.06$ \\
\hline \multicolumn{13}{l}{The columns of Table~3 give: (1)---status the galaxies; (2)---number of galaxies corresponding to
columns (3)--(5); } \\
\multicolumn{13}{l}{(3)---absolute $K$-band magnitudes corrected for extinction according to Melnyk et al. (2017); } \\
\multicolumn{13}{l}{(4)---logarithmic stellar masses (in the units of the solar mass); (5)---absolute $B$-band magnitudes adopted}\\
\multicolumn{13}{l}{from HyperLEDA database corrected for extinction; (6)---the number of galaxies corresponding to
columns (7) and (8);}\\
\multicolumn{13}{l}{(7)---logarithmic star-formation rates $SFR$
(in the units of $M_{\odot}$\,yr$^{-1}$);
(8)---logarithmic specific star-formation }\\
\multicolumn{13}{l}{rates $sSFR$ (in the units of yr$^{-1}$);
(9)---the number of galaxies corresponding to
column~(10);}\\
\multicolumn{13}{l}{(10)---logarithmic H\,I masses $M_{\rm
H\,I}$ (in the units of the solar mass);
(11)---the number of galaxies }\\
\multicolumn{13}{l}{corresponding to columns (12) and (13); (12),
(13)---galaxy colors from the SDSS survey. }
\end{tabular}
\end{table*}
\end{turnpage}

The small size of the sample prevents finding significant
differences between E- and S0-type objects (the two top rows in
Table~3). An expected tendency is immediately apparent with
S0-type galaxies being somewhat bluer than E-type galaxies.

We would like to point out the most peculiar galaxy, KIG\,889,
which we classify as neither elliptical or lenticular, but which
is of interest for a detailed study. We show its PanSTARRS-1 image
in Fig.~1. The size of the field is $100\arcsec\times
100\arcsec$, North is at the top and East is on the left. This
object may be a galaxy with what is well known as conspicuous
``X-shaped structure''. \citet{sav2017} performed detailed
photometry for 22 such objects seen edge-on. A comparison of the
results of simulations demonstrates their qualitative agreement
with observations and supports the  ``bar-driven'' scenario of the
formation of  \mbox {X-shaped}-structures.

 \begin{table*}
\renewcommand{\baselinestretch}{0.8}
\caption {Properties of KIG galaxies and their nearest neighbors
for the determination of orbital masses of isolated galaxies}
\medskip
\begin{tabular}{c|c|c|c|c}
\hline
\multirow{2}{*}{KIG} & \multirow{2}{*}{$M_{K}$} & $dM_{12}$, & $dV$,         & $R_{p}$, \\
                     &                          & mag        & km\,s$^{-1}$  & kpc  \\
\hline
(1) & (2) & (3) & (4) & (5) \\
\hline
228  & $-24.21$ & 3.0 & $271 $ & 163 \\
264  & $-24.07$ & 1.7 & $-110$ & 307 \\
303  & $-24.23$ & 3.1 & $46  $ & 194 \\
303  & $-24.23$ & 1.9 & $130 $ & 306 \\
396  & $-22.60$ & 3.8 & $44  $ & 217 \\
413  & $-23.17$ & 1.2 & $39  $ & 289 \\
413  & $-23.17$ & 2.5 & $278 $ & 306 \\
437  & $-24.62$ & 2.4 & $102 $ & 178 \\
480  & $-23.21$ & 3.4 & $128 $ & 208 \\
517  & $-24.16$ & 3.1 & $62  $ & 140 \\
557  & $-25.05$ & 2.3 & $-198$ & 215 \\
557  & $-25.05$ & 1.7 & $- 72$ & 222 \\
557  & $-25.05$ & 3.7 & $237 $ & 261 \\
578  & $-24.31$ & 3.1 & $7   $ & 161 \\
595  & $-24.94$ & 1.9 & $-174$ & 43 \\
595  & $-24.94$ & 3.1 & $-385$ & 56 \\
596  & $-23.82$ & 1.2 & $118 $ & 209 \\
602  & $-25.04$ & 2.7 & $-136$ & 322 \\
703  & $-23.09$ & 2.6 & $40  $ & 191 \\
722  & $-25.42$ & 3.8 & $122 $ & 186 \\
768  & $-23.40$ & 1.9 & $- 2 $ & 302 \\
771  & $-24.54$ & 2.8 & $-247$ & 13 \\
771  & $-24.54$ & 2.8 & $218 $ & 118 \\
771  & $-24.54$ & 2.0 & $47  $ & 261 \\
771  & $-24.54$ & 2.7 & $-261$ & 288 \\
1042 & $-24.52$ & 1.3 & $-420$ & 261 \\
\hline
Mean & $-24.25\pm0.14$ & $2.53\pm0.15$ & $-5\pm30$ & $208\pm17$ \\
\hline
\end{tabular}
\end{table*}
\renewcommand{\baselinestretch}{1.0}

\begin{figure*}[]
  \includegraphics[scale=0.8]{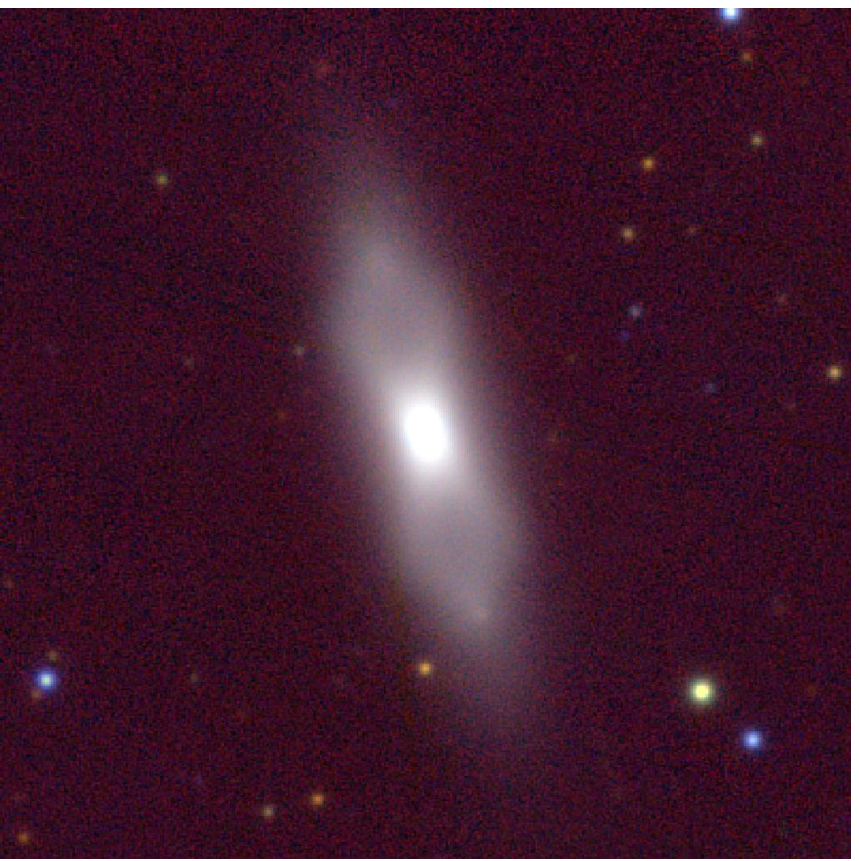}
\caption{Propeller-shaped peculiar galaxy KIG\,889 = NGC\,6969.
The image is taken from PanSTARRS-1. The field has the size of
$100\arcsec\times 100\arcsec$, North is at the top and East is
on the left.}
 \label{Karachentseva_fig1}
\end{figure*}

\section{RESULTS OF A SEARCH FOR SATELLITES/NEIGHBORS AND DESCRIPTION OF THEIR PROPERTIES}

We found in the NED database 112 satellites for 47 isolated
galaxies within the radial velocity difference $\mid dV\mid =
500$~km\,s$^{-1}$ and projected separation $R_{p} = 750$~kpc between
the satellite and KIG galaxy. Two galaxies ---  KIG\,555 and
KIG\,556 --- have radial velocities on the order of
1000~km\,s$^{-1}$ (and  4 and 18 satellites, respectively); we
exclude them from consideration because they reside at the
periphery of Virgo cluster.

The 46+45  isolated galaxies have a total of  90
satellites/neighbors,  i.e., there is about one companion for
every isolated galaxy. This number is about three times less than
\citet{mad2004} obtained for isolated E-type galaxies with $V\leq
2000$~km\,s$^{-1}$. The ratio of the number of satellites to that
of isolated galaxies is higher within the closer volume because of
selection effect \citep{hab2020}. \citet{arg2014} analyzed 386 isolated
KIG galaxies without subdividing them into early- and late-type
systems. A total of 340 (88\%) of these galaxies have no
physically bound satellites. The remaining  46 galaxies have one
to three satellites. We compare the data from our Tables~1 and~2
with Table~1 by \citet{arg2014} and found that there are total of 12
galaxies common with ETG galaxies without satellites and common 27
galaxies with  ETG galaxies having satellites. Of these  11/12
(92\%) are listed in our Table~1 and 13/27 (48\%), in our Table~2.
We can conclude that the results of the comparison are quite good
given different approaches to finding satellites.

Fig.~2 shows the distribution of isolated early-type galaxies
$N_{\rm KIG}$ by the number of satellites.

\begin{figure*}[]
 \includegraphics[scale=0.2]{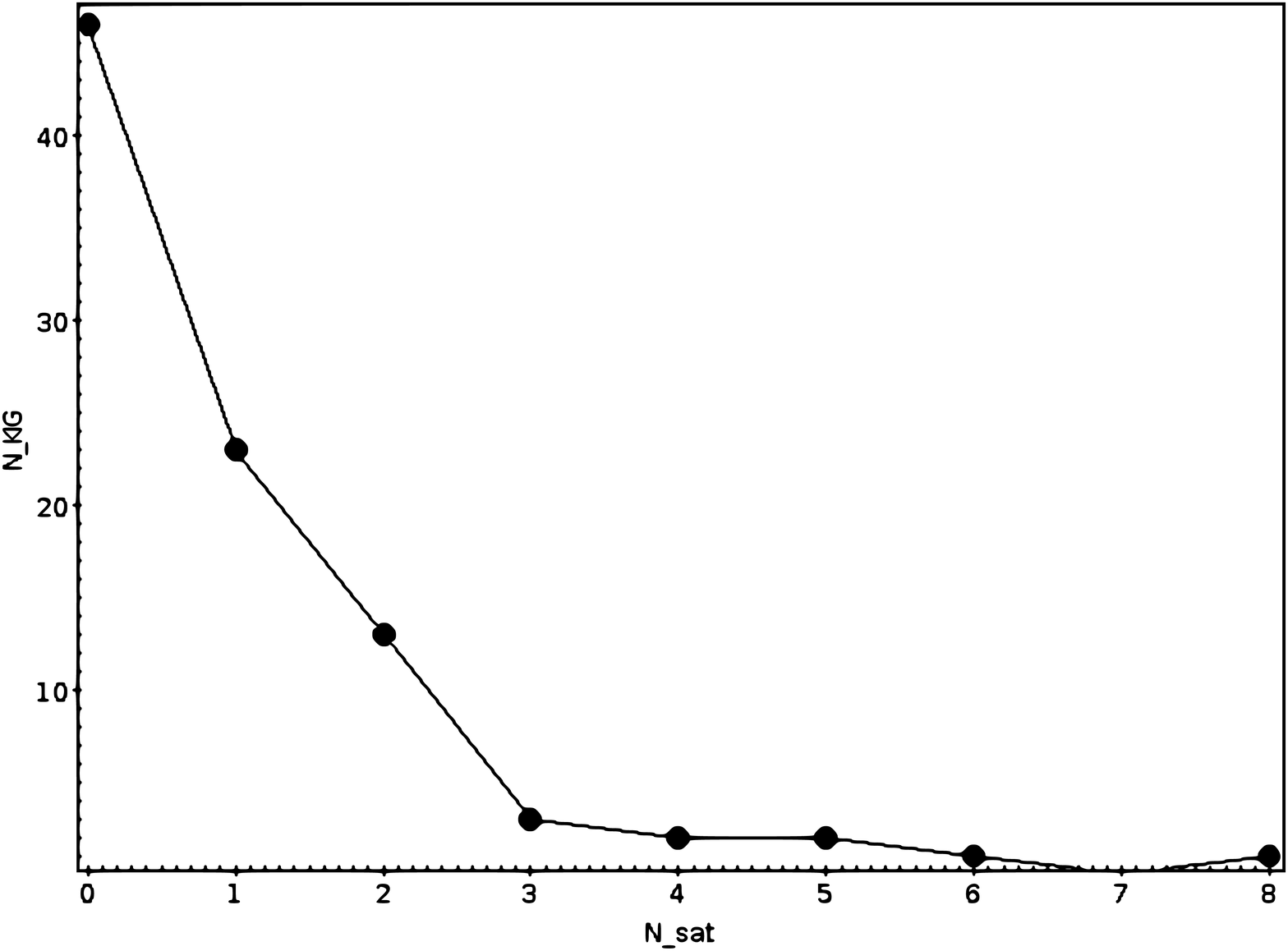}
\caption{Number of isolated ETG galaxies $N_{\rm KIG}$ as a
function of the number of satellites/neighbors $N_{\rm sat}$.}
 \label{Karachentseva_fig2}
\end{figure*}

It follows from Table~2 that satellites and neighbors of isolated
galaxies have morphological type estimates ranging from elliptical
to irregular. The distribution of their types sharply differs with
a greater fraction of both later-type systems and systems with
stronger emission lines) among satellites than among neighbors,
namely:
\begin{list}{}{
\setlength\leftmargin{6mm} \setlength\topsep{2mm}
\setlength\parsep{0mm} \setlength\itemsep{2mm} }
 \item[$\bullet$] satellites: E/S0---29\%; S0a/Sc---23\%; Scd/Sdm---14\%; Sm/Ir---17\%; BCD---17\%;
 \item[$\bullet$] neighbors:   E/S0---20\%; S0a/Sc---55\%; Scd/Sdm---20\%;
Sm/Ir---0\%: BCD---5\%.
\end{list}

The last four rows of Table~3 list the average properties and the
corresponding standard errors for satellites (the last two rows)
and neighbors of isolated galaxies. Although in some cases the
small sample size makes it impossible to draw a definitive
conclusion, certain trends show up: neighbors are significantly
brighter and more massive than satellites, and have greater gas amount
(which is evident from the criteria used to separate them).
Satellites, on the other hand, have somewhat higher star-formation
rates and, on the average, are bluer than neighbors.

Fig.~3 shows the distribution of the absolute values of
radial velocity differences and projected separations between the
KIG galaxies and their satellites, $\vert dV \vert$, km\,s$^{-1}$,
and $R_{p}$, kpc. Satellites and neighbors are shown in the inset
using different symbols.

\begin{figure*}[]
  \includegraphics[scale=0.17]{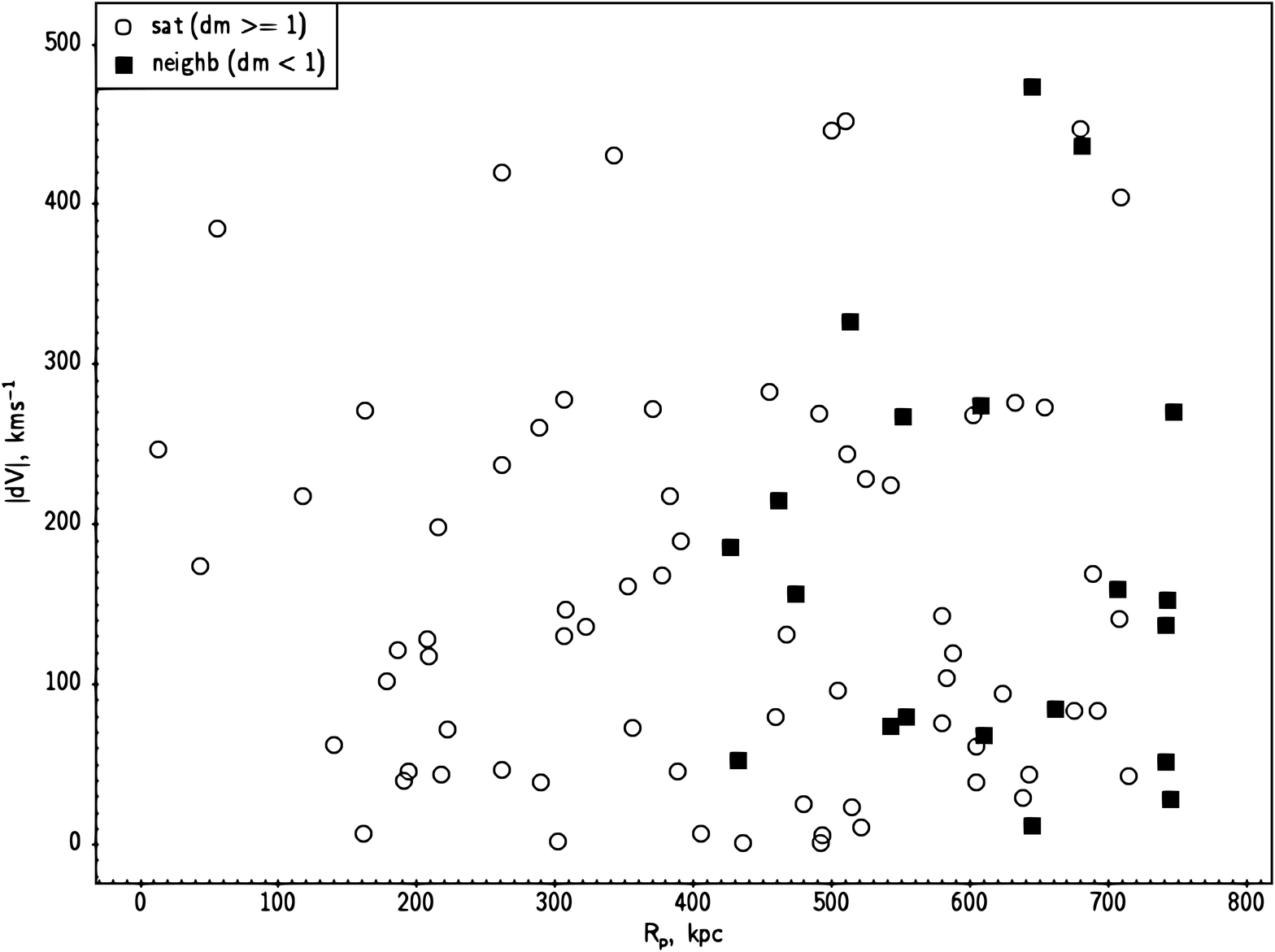}
\caption{Distribution of the radial velocity differences moduli $\vert dV \vert$, km\,s$^{-1}$ and
projected separations $R_{p}$, kpc, between the KIG galaxies and
their companions. The designations are shown in the inset.}
 \label{Karachentseva_fig3}
\end{figure*}

Neighbors, on the average, are located farther than satellites.
One can assume that neighbor galaxies are not gravitationally
bound to KIG galaxies, but belong to a common cosmic filament-like
structure.

Fig.~4 shows the specific star-formation  rate plotted as a
function of stellar mass separately for satellites, neighbors, and
isolated galaxies. Only the upper limit for the $FUV$ flux is
known for about 40\% of KIG galaxies. (We do not show separately
the results of $\log(sSFR)$ computations for these systems in the
figure.) We determine the masses of galaxies from their $K$-band
luminosities assuming that \mbox{$M^*/L_K = 1
M_{\odot}/L_{\odot}$} \citep{bel2003}.

\begin{figure*}[]
  \includegraphics[scale=0.2]{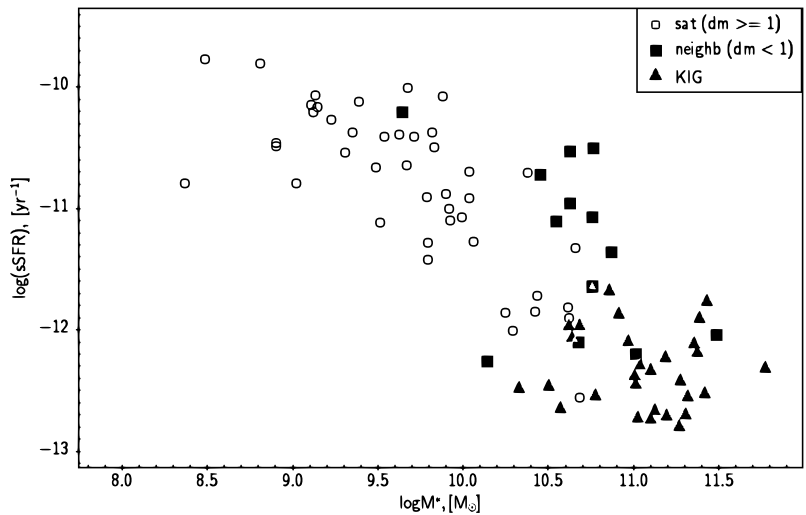}
\caption{Dependence of specific  star-formation rate
$sSFR$, on stellar mass $M^{*}$. The
designations of galaxies are shown in the inset.}
 \label{Karachentseva_fig4}
\end{figure*}

As expected, early-type galaxies in the KIG have quenched star
formation, about the same as we obtained for isolated early-type
galaxies in the 2MIG catalog (see~\citet{mel2015}, Table~1).
Satellite galaxies show a weak decrease of star-formation rate
with increasing stellar mass, whereas neighbor galaxies, whose
magnitudes are approximately equal to those of  ``host'' galaxies,
occupy an intermediate locus in the distribution in Fig.~4 (see
also Table~3).

\section{COMPARISON OF THE PROPERTIES OF EARLY-TYPE KIG GALAXIES WITH AND WITHOUT SATELLITES/NEIGHBORS}

Fig.~5 shows the distribution of radial velocities $V_{\rm
LG}$ of isolated early-type galaxies: (a)~galaxies without
satellites; (b)~galaxies with satellites/neighbors with velocity
differences $\vert dV \vert < 500$~km\,s$^{-1}$ and projected
separations \mbox {$R_{p} <750$}~kpc with  respect to the ``host''
galaxy. The fact that the mean values of the radial velocity
distributions shown in panels (a) and (b) ($8580\pm560$ and
$8184\pm570$~km\,s$^{-1}$, respectively) indicate that KIG
galaxies of both subclasses occupy about the same volume within
the quoted errors with galaxies without satellites being, on the
average, somewhat more distant because of only one outlier galaxy
KIG\,701 with $V_{\rm LG}= 24227$~km\,s$^{-1}$. Note that isolated
early-type galaxies have significantly greater average radial
velocity than all KIG galaxies whose mean radial velocity is \mbox
{$\langle V_{\rm LG}\rangle= 6624$}~km\,s$^{-1}$ according to
\citet{ver2007a}.

\begin{figure*}[]
 \includegraphics[scale=0.2]{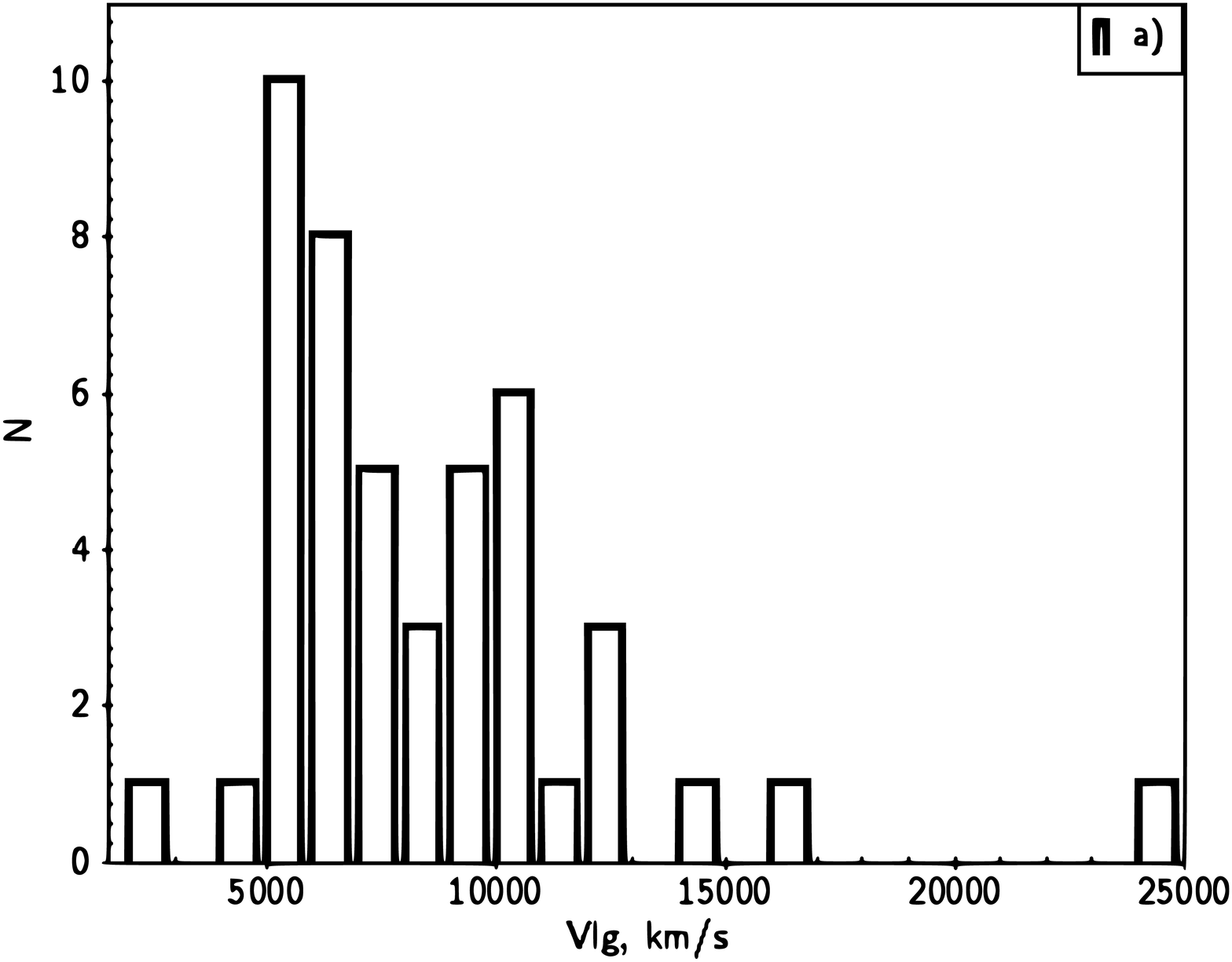}
\includegraphics[scale=0.2]{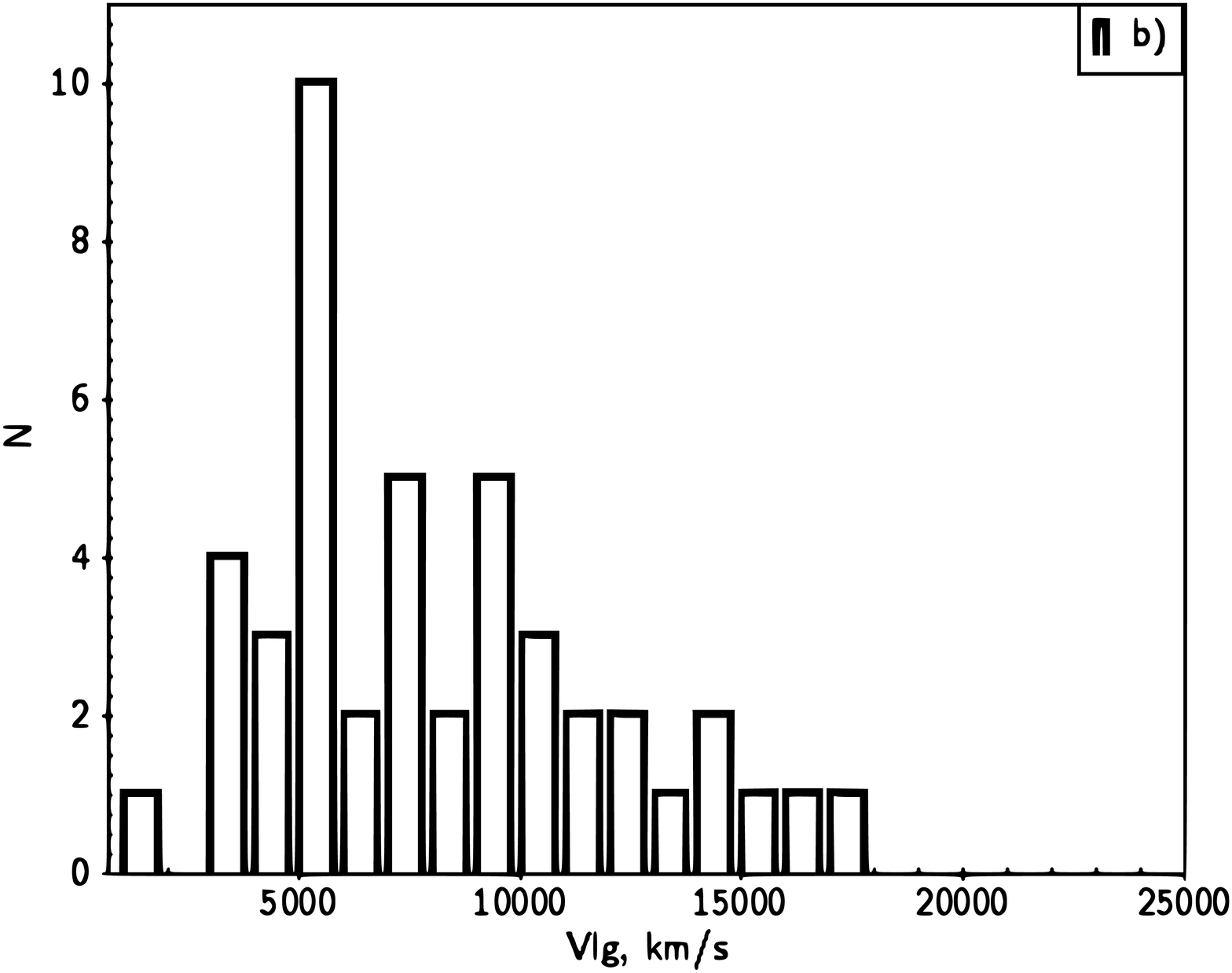}
\caption{Distribution of radial velocities   $V_{\rm LG}$,
km\,s$^{-1}$ of isolated early-type galaxies; (a)--- galaxies
without satellites and (b)---galaxies with satellites having
radial velocity differences $|dV| < 500$~km\,s$^{-1}$ and
projected separations \mbox{$R_{p}<750$}~kpc with respect to the
``host'' galaxy.} \label{Karachentseva_fig5}
\end{figure*}

It follows from the data in Table~3 (rows 3--6) that the mean
absolute magnitudes, specific star formation rates, hydrogen
masses, and colors of isolated E-galaxies and S0-type galaxies do not
differ within the quoted errors.

\section{DETERMINATION OF ORBITAL MASSES OF ETG KIG GALAXIES FROM MOTIONS OF THEIR SATELLITES}

After cleaning the sample of isolated galaxies by types and
excluding two galaxies located in the neighborhood of  Virgo
cluster with $V_{\rm LG} \sim 1000$~km\,s$^{-1}$, there remain a
total of 90 satellites/neighbors with absolute values of the
``satellite--KIG galaxy'' radial velocity differences $\vert dV
\vert < 500$~km\,s$^{-1}$ and projected separations $R_{p} <
750$~kpc. Fig.~3 shows their distributions in the $\vert dV \vert$,
$R_{p}$ plane. As is evident from the figure, at $R_p > 400$~kpc
neighbors appear that are comparable in brightness with KIG
galaxies. Such cases are hardly of any use for estimating the
orbital masses. On the other hand, the projected virial halo
radius for our Milky Way galaxy and  M\,31 with their $K$-band
luminosities $L_{K} \sim 5\times 10^{10}L_{\odot}$ is of about
250~kpc~\citep{tul2015}. The average luminosity of an ETG galaxy
with satellites from Table~2 is \mbox {$L_{K} \sim 1.0\times
10^{11} L_{\odot}$}, i.e., twice higher. Given that the mass of
the halo is proportional to the cube of the virial radius, the
virial radius of a typical KIG ETG galaxy may be as large as about
330~kpc. Therefore hereafter we consider only the KIG galaxies
with satellites with $R_{p} < 330$~kpc, and there a total of 26
such cases. We summarize the results in Table~4. Its columns give:
(1)---the number of the galaxy in the KIG catalog; (2)---corrected absolute
 \mbox {$K_s$-band} magnitude from NED; (3)---the difference of the
absolute $K$-magnitudes between the
satellite and KIG galaxy; (4)---the difference of radial
velocity between the satellite and the KIG galaxy in km\,s$^{-1}$;
(5)---the mutual projected separation $R_{p}$ in kpc. The last row
gives the average parameter values and their standard error.

The data from Table~4 lead us to conclude that:
\begin{list}{$\bullet$}{
\setlength\leftmargin{6mm} \setlength\topsep{2mm}
\setlength\parsep{0mm} \setlength\itemsep{2mm} }
 \item a ``typical'' satellite is ten times fainter than its KIG host galaxy, i.e., for these bound
systems the Keplerian approach can be used to determine the mass
of the central dominating galaxy from the motions of its small
satellites;
 \item   the mean difference of the radial velocities of satellites is close to zero,
$\langle dV\rangle = -5\pm30$~km\,s$^{-1}$, and this fact supports
their physical connection with the corresponding KIG galaxies;
 \item   at
$\langle M_{K}\rangle = -24.25\pm0.14$ a KIG ETG galaxy has a
luminosity of $\log(L_K) = 11.01\pm0.06$, or $L_{K} =
(1.03\pm0.15) \times 10^{11}L_{\odot}$, which is twice higher than
the luminosity of the Milky Way.
\end{list}

Under the assumption of random orientation of satellite orbits
with a mean orbital eccentricity of $\langle e \rangle = 0.7$
\citep{bar2014} the mass of the central object can be written as
$M_{\rm orb} = (16/ \pi G) \langle dV^{2}R_{p}\rangle$, where $G$
is the gravitational constant. Based on the data for 26 satellites
listed in Table~4 we estimated  the orbital mass as $$M_{\rm
orb} = (7.56 \pm 2.36)\times 10^{12}~M_\odot,$$  i.e. a halo
mass to the average \mbox {$K$-band} luminosity ratio for  E- and
S0-type KIG galaxies of $$M_{\rm orb}/L_K = 74 \pm 26.$$

This ratio is close to the corresponding ratios $M_{\rm
orb}/L_{K}$~=~$38\pm22$, $82\pm26$, and $65\pm20$ for the massive
Local-Volume ETG galaxies  NGC\,3115, NGC\,5128, and NGC\,4594,
respectively \citep{kar+kud2014, kar2020}. At the same time, the
average orbital mass to $K$-band luminosity ratio for apparently
bulgeless spiral galaxies is as low as
$(20\pm3)M_{\odot}/L_{\odot}$ \citep{kar+kara2019}.

\citet{kara2011} analyzed the velocities and projected separations
of dwarf satellites located in the vicinity of 2MIG galaxies and
found that the motions of 60 satellites about E- and S0-type
galaxies imply a median ratio of $M_{\rm orb}/L_{K}  = 63$,
whereas the data for 154 satellites orbiting spiral galaxies yield
a median ratio of $M_{\rm orb}/L_{K} = 17$.  This about threefold
difference between the dark-to-visible mass ratios is an
indication suggesting that the dynamic evolution of early- and
late-type galaxies proceeded along essentially different
scenarios.

\section{CONCLUDING REMARKS}

Isolated early-type  (E, S0) galaxies and galaxies of the same
types residing in groups and clusters may have different dynamic
history and structure. A standard sample of elliptical and
lenticular galaxies is needed to reveal such differences. In this
study we use the Catalog of Isolated Galaxies (KIG,
\citep{kara1973}) as such standard sample. It contains 1050
objects, which makes up for about 4\% of all Northern-hemisphere
galaxies with apparent magnitudes $m_{B}\leq 15.7$~mag. Of these
only165 galaxies were classified as belonging to types E and S0.
Hence isolated early-type galaxies are a rather rare  (0.6\%)
category of galaxies in the \citet{Zwi1968} catalog. The small
number of such galaxies is consistent with the idea that  E- and
S0-type galaxies form as a result of mergers or close interactions
of neighbors.

We use modern digital sky surveys (PanSTARRS-1, SDSS) combined
with the data of  H\,I-line and far-ultraviolet (GALEX) sky
surveys to reclassify 165 early-type galaxies in the KIG. As a
result, the number of E- and S0-type galaxies was reduced down
to~91. Our classification of these galaxies and the classification
performed by other authors are presented in Tables~1 and~2. About
20\% of the galaxies of this sample exhibit various peculiarity
features (anomalous structure, emissions in optical lines,
presence of H\,I or $FUV$ fluxes).

Lenticular and elliptical galaxies have, on the average, high
$K$-band luminosities:
$$\langle \log L_{K(S0)}\rangle =
11.02\pm0.04$$ and  $$\langle \log L_{K(E)}\rangle =
10.99\pm0.05$$ in the solar units. Note that S0-type galaxies
appear somewhat bluer $$\langle g-r \rangle = 0.75\pm 0.02,~~~
\langle g-i \rangle = 1.10 \pm 0.05,$$  compared to E galaxies
with $$\langle g-r \rangle = 0.81 \pm0.01,~~~ \langle g-i \rangle
= 1.21 \pm 0.01.$$

Our search for satellites of early-type KIG galaxies revealed 90
neighbors with radial velocity differences $\vert dV \vert
<500$~km\,s$^{-1}$ and linear projected separations $R_{p} <
750$~kpc. Note that half of KIG galaxies have no neighbors with
such properties.

We found no appreciable differences in either integrated
luminosities or colors of ETG KIG galaxies due to the presence or
absence of close neighbors.

An average early-type KIG galaxy is twice more luminous that the
Milky Way or M\,31 and has a characteristic virial radius of about
330~kpc. There are 26 satellites within this radius and their
average luminosity is one order of magnitude lower than that of
KIG galaxies. The presence of such small satellites does not contradict
 the isolation criterion adopted in the KIG.

We assumed that the orbits of 26 satellites are randomly oriented
and that their average eccentricity is equal to  \mbox {$\langle e
\rangle = 0.7$} to infer the average orbital mass of E- and
S0-type KIG galaxies, which we found to be $$M_{\rm orb} =
(7.56\pm 2.36)\times 10^{12} M_{\odot}.$$ The
characteristic orbital mass to luminosity ratio of isolated  E-
and S0-type galaxies
$$M_{\rm orb}/L_{K} = (74\pm26) M_{\odot}/L_{\odot}$$ is consistent
with the  $M_{\rm orb}/L_K$ estimates
for isolated early-type galaxies in the  2MIG catalog ($63
M_{\odot}/L_{\odot}$), as well as with the $M_{\rm orb}/L_{K}$
estimates for E- and S0-type galaxies in the Local Volume: \mbox
{$38\pm22$} (NGC\,3115), $82\pm26$ (NGC\,5128), and $65\pm20$
\linebreak (NGC\,4594) in the solar units.

The high halo mass to luminosity ratio for E- and S0-type galaxies
compared to the corresponding average ratio $(20\pm3)
M_{\odot}/L_{\odot}$ for bulgeless spiral galaxies is indicative
of essential differences between the dynamic evolution of early-
and late-type galaxies.

\section*{ACKNOWLEDGMENTS}
This work made use of  PanSTARRS1, SDSS, 2MASS, and GALEX sky
surveys and  HyperLEDA (http://leda.univ-lyon1.fr) and NED
\linebreak (http://ned.ipac.caltech.edu/) databases.

\section*{FUNDING}
This work was supported in part by the program of the National
Academy of Sciences of Ukraine  (CPCEL 6541230). IDK acknowledges
financial support from the Russian Science Foundation (grant no.
\mbox {19-12-00145}).

\section*{CONFLICT OF INTEREST}
The authors declare no conflict of interest.
%

{}

\end{document}